\documentclass[11pt,a4paper]{article}
\usepackage{jheppub}

\usepackage{color}
\usepackage{amsfonts}
\usepackage{slashed}

\def\Det{{\rm Det}}
\def\tr{{\rm Tr}}

\def\bea{\begin{eqnarray}}
\def\eea{\end{eqnarray}}

\def\lmatrix{\left(\begin{array}}
\def\rmatrix{\end{array}\right)}

\def\msbar{\overline{\rm MS\kern-0.5pt}\kern0.5pt}
\def\leqx{\,\raisebox{-1.0ex}{$\stackrel{\textstyle <}{\sim}$}\,}

\title{The running coupling of the minimal sextet composite Higgs model}

\author[abc]{Zoltan Fodor,}
\author[de]{Kieran Holland,}
\author[f]{Julius Kuti,}
\author[cg]{Santanu Mondal,}
\author[cgh]{Daniel Nogradi}
\author[a]{and Chik Him Wong}

\affiliation[a]{University of Wuppertal, Department of Physics, Wuppertal D-42097, Germany}
\affiliation[b]{J\"ulich Supercomputing Center, Forschungszentrum J\"ulich, J\"ulich D-52425, Germany}
\affiliation[c]{E\"otv\"os University, Institute for Theoretical Physics, Budapest 1117, Hungary}
\affiliation[d]{University of the Pacific, 3601 Pacific Ave, Stockton CA 95211, USA}
\affiliation[e]{Albert Einstein Center for Fundamental Physics, Institute for Theoretical
Physics, Bern University, Sidlerstrasse 5, CH-3012 Bern, Switzerland}
\affiliation[f]{University of California, San Diego, 9500 Gilman Drive, La Jolla, CA 92093, USA}
\affiliation[g]{MTA-ELTE Lendulet Lattice Gauge Theory Research Group, 1117 Budapest, Hungary}
\affiliation[h]{Kavli Institute for Theoretical Physics, University of California Santa Barbara, CA 93106, USA}

\emailAdd{fodor@bodri.elte.hu}
\emailAdd{kholland@pacific.edu}
\emailAdd{jkuti@ucsd.edu}
\emailAdd{santanu@bodri.elte.hu}
\emailAdd{nogradi@bodri.elte.hu}
\emailAdd{cwong@uni-wuppertal.de}

\preprint{NSF-KITP-15-126}

\abstract{ 
We compute the renormalized running coupling of $SU(3)$ gauge theory coupled to $N_f = 2$ flavors of massless
Dirac fermions in the 2-index-symmetric (sextet) representation. This model is of particular interest as a minimal
realization of the strongly interacting composite Higgs scenario.
A recently proposed finite volume gradient flow scheme is used. The calculations are
performed at several lattice spacings with two different
  implementations of the gradient flow  allowing for a controlled continuum 
extrapolation and particular attention is
paid to estimating the systematic uncertainties. For small values of the
renormalized coupling our results for the $\beta$-function agree with perturbation theory. For moderate couplings
we observe a downward deviation relative to the 2-loop $\beta$-function but in the coupling
range where the continuum extrapolation is fully under control we do not observe an infrared fixed point. 
The explored range includes the locations of the zero of the 3-loop and the 4-loop $\beta$-functions in the $\msbar$
scheme. The absence of a non-trivial zero in the $\beta$-function in the explored range of the coupling is consistent with our
earlier findings based on hadronic observables, the chiral condensate and the GMOR relation. 
The present work is the first to report continuum non-perturbative results for the sextet model.
}


\arxivnumber{1506.06599}

\begin{document}

\maketitle

\section{Introduction}

We study $SU(3)$ gauge theory coupled to $N_f = 2$ flavors of massless Dirac fermions in the 2-index-symmetric (sextet)
representation. The model may be a minimal realization of the strongly interacting composite Higgs scenario 
\cite{Sannino:2004qp,Hong:2004td,Dietrich:2005jn,Dietrich:2006cm, Fodor:2012ty, Fodor:2012ni,
Fodor:2014pqa, Fodor:2015vwa} 
which is our primary motivation. This possibility has attracted active experimental interest recently \cite{Aad:2015yza}. 
Since the dynamics is intrinsically strongly coupled only a non-perturbative approach like the
lattice can determine the ultimate viability of the model. Several simulation results have been reported, but none which fully controls all
systematic errors and provides a consistent picture of the various aspects of the model such as the running coupling, the anomalous mass
dimension, finite temperature, hadronic observables, etc. Recent reviews are given in \cite{Nogradi:2012rq, Fodor:2015vwa}.

Even though the currently available simulations are performed at finite lattice spacing and may be affected by other uncontrolled
systematics, the evidence suggested by these results is that the infrared behavior of the model is QCD-like. Chiral
symmetry is broken spontaneously~\cite{Fodor:2011tw, Fodor:2012uu, Fodor:2012ty, Fodor:2012uw, Fodor:2015vwa}
leading to a thermal phase transition at finite temperature
\cite{Sinclair:2009ec, Sinclair:2010be, Kogut:2010cz, Sinclair:2011ie, Kogut:2011ty, Sinclair:2012fa}.
Results on rough lattices concerning the running coupling in the Schroedinger-functional scheme 
can be found in \cite{Shamir:2008pb, DeGrand:2010na}.

The goal of the present work is to calculate the $\beta$-function of the renormalized running coupling over a wide range
of couplings in such a way that all systematics are fully controlled. We are interested in the massless theory and
our simulations are set up in such a way that the bare mass can be set to zero directly. The running is implemented via
changing the physical volume of the system, introducing a scale $\mu = 1/L$, 
hence finite volume dependence is translated into the renormalization group flow of the coupling. This leaves
us with the continuum extrapolation as the only source of systematic uncertainties.

By performing the simulations at several pairs of lattice volumes
in the framework of a step scaling analysis, we gain control over the finite lattice spacing effects. 
In fact at each value of the renormalized coupling, results at five values of the lattice spacing are used, with two different discretizations for the observables in question and
a reliable continuum extrapolation is hence possible. Our work is the first to report on fully controlled continuum
results for the $N_f = 2$ sextet model.

\section{The gradient flow running coupling scheme}
\label{thegradientflow}

The gradient flow \cite{Morningstar:2003gk, Narayanan:2006rf, Luscher:2009eq, Luscher:2010iy, Luscher:2010we,
Luscher:2011bx, Lohmayer:2011si} is a particularly useful 
tool for studying the running coupling of a non-abelian gauge
theory. There are various finite volume setups which mainly differ in the choice of boundary conditions for the gauge
field
\cite{Fodor:2012td, Fodor:2012qh, Fritzsch:2013je, Fritzsch:2013hda, Ramos:2013gda, Rantaharju:2013bva, Luscher:2014kea, Ramos:2014kla}.
In the present work we follow \cite{Fodor:2012td, Fodor:2012qh} where the gauge field is taken as periodic in all four
directions. For the fermions on the other hand we impose anti-periodic boundary conditions again in all four directions.
Other applications of the gradient flow can be found in 
\cite{Borsanyi:2012zs, Asakawa:2013laa, Suzuki:2013gza}.

More precisely, in our scheme a 1-parameter family of couplings is defined in finite 4-volume $L^4$ by
\bea
\label{g}
g_c^2 = \frac{128\pi^2\langle t^2 E(t) \rangle}{3(N^2-1)(1+\delta(c))}\;,\qquad E(t) = -\frac{1}{2} \tr F_{\mu\nu} F_{\mu\nu}(t)
\eea
where $t$ is the flow parameter, $N$ corresponds to the gauge group $SU(N)$, $c = \sqrt{8t}/L$ is a constant, $E(t)$ is
the field strength squared at $t > 0$ and the numerical factor
\bea
\label{delta}
\delta(c) = - \frac{c^4 \pi^2}{3} + \vartheta^4\left(e^{-1/c^2}\right) - 1
\eea
is chosen such that at leading order $g_c^2$ agrees with the coupling in $\msbar$ for all $c$; $\vartheta$ is the 3rd
Jacobi elliptic function.
Hence the coupling $g_c(\mu)$ runs via the scale $\mu = 1/L$, for more details see \cite{Fodor:2012td, Fodor:2012qh}.

There is a peculiarity of our running coupling scheme related to the fact that we 
impose periodicity on the gauge fields, leading to zero modes
\cite{Luscher:1982ma,Koller:1985mb,Koller:1987fq,vanBaal:1988va,vanBaal:1988qm,
KorthalsAltes:1985tv,Coste:1985mn,Coste:1986cb,KorthalsAltes:1988is}. These gauge zero modes cause the perturbative
expansion of $g_c^2$ in $g_{\msbar}$ to contain both even and odd powers and potentially logarithms too. 
However for $N>2$ logarithms do not appear in the first two non-trivial orders, only polynomials
\cite{Fodor:2012qh}. The first unusual odd power
results in only the 1-loop $\beta$-function coefficient being the same as in $\msbar$.

The constant $0 < c \leq 1/2$ specifying the scheme can in principle be chosen at will. However, as discussed in
\cite{Fodor:2012td, Fodor:2012qh}, a small $c$ leads to small statistical errors but large cut-off effects and a larger
$c$ results in larger statistical errors and smaller cut-off effects. In the present work we set $c=7/20$, slightly
higher than the value $c=3/10$ in \cite{Fodor:2012td, Fodor:2015baa}, in order to reduce cut-off effects.

\section{Rooted staggered formulation}
\label{rootedstaggeredformulation}

The fermion doublet in the staggered fermion implementation requires the square root of the fermion 
determinant, also known as the
 rooting procedure.  With the mass of the fermion doublet set to zero, the continuum 
 step $\beta$-function as determined from a scale-dependent renormalized coupling $g_R^2 (L)$ shows no sign of
 turning zero in the explored range, as we will see.  Our results are consistent with chiral symmetry breaking
 when probed with finite fermion mass deformations in the p-regime ~\cite{Fodor:2015vwa}.
  
Some preliminary work, with goals similar to ours but in the Wilson fermion formulation, reports consistency with 
a zero in the $\beta$-function in the renormalized running coupling
in the same range where our $\beta$-function is positive and monotonically growing~\cite{annatalks2, annatalks1, annatalks3}.
If confirmed and 
further supported with conformal scaling laws, the new work in Wilson formulation might suggest a conformal 
infrared fixed point at vanishing fermion mass in the sextet  model, 
inconsistent with our results and spontaneously broken chiral symmetry.

Motivated by this controversy, doubts were raised about our results questioning the application of the rooting procedure 
with the mass of the staggered fermion doublet set to zero 
in the simulations at finite lattice spacing. This lead to the speculation
that in the staggered rooting procedure setting the fermion mass $m$ to zero at fixed finite lattice spacing $a$ 
might be incorrect because of the 
non-locality of the rooted staggered action we appear to deploy by interchanging the so-called required limit of 
$a \rightarrow 0$ first, while holding the fermion mass 
non-zero before taking the $m\rightarrow 0$ limit in the continuum theory as the last step, after the cutoff is removed.
As a consequence of this issue of  non-locality concerns were raised whether a rooted staggered theory is in the
correct universality class of the continuum theory and whether rooting can identify a conformal theory~\cite{annatalks2, annatalks1, annatalks3}.

To alleviate the concerns, we will show that the rooting procedure is correct when 
the fermion mass is set to zero at finite lattice spacing while the finite physical volume of the continuum limit is held fixed.
Consequently, the conformality of a model would not be missed and
the rooted staggered formulation in finite physical volume and in the  infinite volume limit are expected to remain in the 
correct universality class.

\subsection{Review of rooting in infinite volume}

The method to address the rooting procedure properly has been developed in a series of papers by Bernard, Golterman, Shamir, 
and Sharpe~\cite{Bernard:2004ab,Bernard:2006zw,Bernard:2006ee,Sharpe:2006re,Shamir:2006nj,Bernard:2006vv,Bernard:2007ma} when the renormalized fermion
mass is kept finite before the continuum limit is taken.
We adapt their analysis to our model in finite physical volumes to demonstrate that the rooting procedure we apply
at vanishing fermion mass and finite lattice spacing $a$ should remain valid 
on the level of their reasoning.

The main results of the analysis at finite fermion mass and infinite volume are summarized first
from two succinct exposures of the rooting issues~\cite{Shamir:2006nj,Bernard:2007ma} that we closely follow here. 
Accordingly, the rooted staggered action is defined on a fine-grained lattice with lattice spacing $a_f$ and connected with infrared physics
using $n$ renormalization group steps to a blocked  physically equivalent lattice action 
on a coarse lattice with lattice spacing $a_c$ which is held fixed on some physical scale~\cite{Shamir:2006nj,Bernard:2007ma}.
At fixed $a_c\Lambda_{IR}$, where $\Lambda_{IR}$ designates some non-perturbative infrared scale, the 
continuum limit $a_f \rightarrow 0$ is 
investigated when $n\rightarrow\infty$.

In the technical implementation of the RG procedure, the blocked and unrooted staggered
Dirac operator $D_{\rm stag,n}$  is split into the taste invariant part $D_{\rm inv,n}=D_n\otimes{\bf 1_4}$  with exact 
taste symmetry of four degenerate fermions  and the taste breaking part $\Delta_n$  after each blocking step, 
\begin{equation}
D_{\rm stag,n}=D_{\rm inv,n} +\Delta_n , ~~~~  D_{n}=\frac{1}{4}\;\tr\left(D_{\rm stag,n}\right) ,
\end{equation}
where $\tr$ denotes the trace in taste space. The trace of $\Delta_n$ vanishes in taste space, 
and ${\bf 1_4}$ designates the taste identity matrix.
The  local taste invariant theory represented by $D_{n}\otimes{\bf 1_4}$ has four degenerate fermions in taste space and the fourth root 
of the fermion  determinant is trivially given by 
\begin{equation}
\Det^{1/4}(D_{n}\otimes{\bf 1_4})=\Det(D_{n})~.
\end{equation}
After taking the fourth root of $D_{\rm stag,n}$, an estimate is needed to show the convergence of 
$\Det^{1/4}(D_{\rm stag,n})$ to the local single taste determinant $\Det (D_{n})$ in the $n\rightarrow\infty$ limit.
As shown in Eq.~(\ref{eq:rooted}), the convergence of this expansion is controlled by $\|D_{\rm inv,n}^{-1}\Delta_n\|$.
We need to estimate $\|a_c\Delta_n\|$ and $\|(a_c D_{\rm inv,n})^{-1}\|$ separately on the scale of the coarse lattice.

Following~\cite{Shamir:2006nj} we can safely assume the bound $\|a_c\Delta_n\|\;\leqx\; a_f/a_c$ to hold 
on the coarse lattice and scaling with $a_f$.
This is a basic feature of unrooted fermions simply stating that taste-breaking disappears in the continuum limit.  
By exploiting the proximity of the local re-weighted theory defined with $D_{\rm inv,n} $
after a large number $n$ of blocking steps, it was argued that the bound $\|a_c\Delta_n\|\;\leqx\; a_f/a_c$ and 
its scaling with $a_f/a_c $ is also valid in rooted theories~\cite{Shamir:2006nj}.
We will adapt this argument. The estimate for the upper bound on the inverse of the taste invariant operator is given by
$\|(a_c D_{\rm inv,n})^{-1}\|  \leq 1/(a_c m_R(a_c))$  with the renormalized fermion mass set at the physical scale $a_c$.
The important combined estimate follows with 
\begin{equation}
\label{eq:estimate}
\|D_{\rm inv,n}^{-1}\Delta_n\| \leq a_f/(a_c^2 m_R(a_c))
\end{equation}
and the small expansion parameter
\begin{equation}
\label{eq:epsilon}
\epsilon_n = \|a_c\Delta_n\|\cdot \|(a_c D_{\rm inv,n})^{-1}\| \leq a_f/(a_c^2 m_R(a_c)) = \frac{1}{2^{n+1}a_c m_R(a_c)}
\end{equation}
where in the first step the lattice spacing is doubled by the change from staggered fermion basis to Dirac basis
followed by n blocking steps in the Dirac basis. This small expansion  parameter implies
the convergence of the rooted staggered theory to a local action of a single taste in the $n\rightarrow\infty$ limit,
\begin{eqnarray}
\label{eq:rooted}
\Det^{1/4}\left(D_n\otimes{\bf 1}_4+\Delta_n\right) \nonumber
&=&\Det(D_n)\;\exp\left[\frac{1}{4}\tr\log\left({\bf 1}_4+D_{\rm inv,n}^{-1}\Delta_{n}\right)\right]\\ 
&=&\Det(D_n)\left(1+O\left(\frac{a_f}{a_c^2m_R(a_c)}\right)\right)  .
\end{eqnarray}
It is important to note that in estimating a lower bound on the norm of $ (a_c D_{\rm inv,n})^{-1}$ the finite renormalized fermion mass $m_R(a_c)$
provides the infrared cutoff of the Dirac spectrum when the volume is infinite. Since renormalization is multiplicative in the staggered formulation, 
it is implemented on the physical scale $a_c$ requiring the adjustment of the bare mass in the $n\rightarrow\infty$ limit which is equivalent to the
$a_f\rightarrow 0$ continuum limit. The choice $m_R(a_c)$ is arbitrary once a physical scale $a_c$ is set in the theory. This is expected because
the RG invariant fermion mass related to $m_R(a_c)$ is arbitrary but as long as it is kept finite in the estimate of the expansion 
in Eq.~(\ref{eq:rooted}) the convergence of the rooted theory to the local single taste action is assured. 

This line of reasoning, based on~\cite{Shamir:2006nj,Bernard:2007ma},  explains why the sextet model of the rooted fermion doublet with
finite renormalized mass is expected to be in the correct universality class when  the continuum limit is taken.
In our simulations of the volume dependent  
renormalized coupling $g_R^2$ the renormalized fermion mass $m_R(a_c)$ is set to zero on any physical scale $a_c$ since the bare mass $m$
itself is set to zero and the mass renormalization is multiplicative from the chiral symmetry of the staggered formulation. Since the estimate of the expansion
for the convergence of the rooted determinant to the local single taste determinant as given in Eq.~(\ref{eq:rooted}) is not applicable in the work
presented here, the rooting procedure at $m_R(a_c)=0$  requires separate discussion.

\subsection{Rooting in finite physical volume at zero bare mass}

It is important to precisely define our rooting procedure where the required limit suggested by  Eq.~(\ref{eq:rooted}) is not followed. 
In all of the simulations reported in this paper the bare fermion mass is set to zero at finite lattice spacing $a$. 
This also sets the renormalized mass 
to zero on any choice of physical scale $a_c$ on the coarse lattice.
Although the estimate on the bound and its scaling with the RG steps in 
Eq.~(\ref{eq:epsilon}) is lost, there is no problem with the rooting procedure since anti-periodic boundary conditions are imposed 
on the fermions in all four directions. 

As we will now show, the choice of anti-periodic boundary conditions for the fermions restores 
the validity of the rooting procedure. The simulations always target some chosen values of the  scale-dependent renormalized coupling. 
Each choice selects the corresponding  linear size $L$ of the physical volume in the continuum. 
The renormalized coupling $g_R^2(a_c)$ in the finite volume $L$ also depends on the ratio $a_c/L$ 
from a 1-parameter family of schemes. 
As the number of RG steps keeps increasing toward the continuum limit, the coupling on the scale $a_f$ 
(bare coupling on the cutoff scale) has to be adjusted while $g_R^2(a_c)$ is held fixed, and similarly the lattice size measured in $a_f$ units 
is adjusted to keep the physical size $L$ fixed together with $a_c/L$. The scheme we introduced in section~\ref{thegradientflow} defines 
a different but related finite volume scheme without affecting the reasoning. The former is built on the RG procedure and the other is defined 
on the gradient flow.

In the finite volume scheme a finite gap  $\lambda_{\rm gap}(a_c)$ is created in the Dirac spectrum which depends on $g_R^2(a_c)$. 
Weak couplings correspond to small physical scales and the gap  is approximately determined by the minimum momentum $\pi/L$ in each direction with ${\cal O}(g_R^2(a_c))$
corrections. As the renormalized coupling become stronger with increasing volume and the interacting energy levels 
increasingly repel, 
they settle into a gradually decreasing but finite gap $\lambda_{\rm gap}(a_c)$ set by the physical scale of the volume.
The estimate  for the bound on the taste breaking operator remains unchanged with $\|a_c\Delta_n\|\;\leqx\; a_f/a_c$.
The new estimate for the upper bound on the inverse of the taste invariant operator is given by
$\|(a_c D_{\rm inv,n})^{-1}\|  \leq 1/(a_c \lambda_{\rm gap}(a_c))$  with the gap of the spectrum set at the physical scale $a_c$.
The important combined estimate follows with $\|D_{\rm inv,n}^{-1}\Delta_n\| \leq a_f/(a_c^2\lambda_{\rm gap}(a_c) ) $ 
and the small expansion parameter is changed to 
\begin{equation}
\label{eq:epsilon1}
\epsilon_n = \|a_c\Delta_n\|\cdot \|(a_c D_{\rm inv,n})^{-1}\| \leq a_f/(a_c^2\lambda_{\rm gap}(a_c) ) = \frac{1}{2^{n+1}a_c\lambda_{\rm gap}(a_c)} .
\end{equation}
The finite gap in the Dirac spectrum implies the convergence of the rooted staggered theory to a local action of a single taste in the $n\rightarrow\infty$ limit,
\begin{eqnarray}
\label{eq:rooted1}
\Det^{1/4}\left(D_n\otimes{\bf 1}_4+\Delta_n\right) \nonumber
&=&\Det(D_n)\;\exp\left[\frac{1}{4}\tr\log\left({\bf 1}_4+D_{\rm inv,n}^{-1}\Delta_{n}\right)\right]\\ 
&=&\Det(D_n)\left(1+O\left(\frac{a_f}{a_c^2\lambda_{\rm gap}(a_c)}\right)\right)  .
\end{eqnarray}
The conclusion from this simple analysis is that in the calculation of the volume dependent running coupling with a rooted and massless fermion doublet the
role of the renormalized mass $m_R(a_c)$ on the physical scale is replaced by the $\lambda_{gap}(a_c)$ gap of the Dirac operator in the finite physical
volume set by the targeted renormalized coupling $g_R^2(a_c)$ for fixed $a_c/L$. It is easy to see how this works at weak coupling and is sustained with growing
volume if the gap does not collapse to zero at some critical volume size. 
At any targeted value of $g_R^2(L)$ while holding the physical size $L$ fixed, the 
eigenvalues of the infrared Dirac spectrum will collapse into
degenerate quartets in the $a_f\rightarrow 0$ limit, consistent with the locality of the rooted action in the continuum limit.

We know, however, that although the simulations become increasingly difficult with
increasing volume, the  ensemble-averaged gap cannot disappear in finite physical volumes not even after some rapid crossover into the phase which either has chiral symmetry breaking or is instead conformal.
With chiral symmetry breaking, the low end of the spectrum is expected to scale as $\lambda \sim 1/V$ which protects the gap.
In the conformal theory the spectral density is expected to scale as $\rho(\lambda) \sim \lambda^\alpha$ with some critical 
exponent
$\alpha$ and  $\lambda  \sim (1/ L)^{4/(1+\alpha)}$ 
for the low infrared part of the spectrum which also protects the gap from complete collapse.
The finite gap  cannot disappear in finite physical volumes even if the rooted model is conformal.   In the conformal case the beta function 
is expected to turn zero at some critical coupling $g_{{\rm crit}}^2$ which can only be reached asymptotically at infinite volume. 
Our method with rooted staggered fermions can clearly distinguish a conformal model from one with chiral symmetry breaking.

The simulations, as reported in this paper, reach a limited range in the renormalized coupling without any sign of the $\beta$ function turning zero. 
Beyond our reach, the results do not rule out conformality in large volumes,  although they remain consistent with chiral symmetry breaking.
Although it is tempting to pursue further simulations in large volumes at vanishing fermion mass, it is not practical at very small values of the
gap in the Dirac spectrum. 

\subsection{The bridge to large volume physics and simulations at finite cutoff  ${\bf a_f}$}

Studying small fermion mass deformations in large volumes at finite cutoff  $a_f$ 
can clearly differentiate between phases with chiral symmetry breaking, or conformality. Taste breaking at finite $a_f$ is
described by operators in the Symanzik effective theory (SET) as calculated in~\cite{Lee:1999zxa,Sharpe:2004is}.
As pointed out in~\cite{Bernard:2007ma}, when the goal is to match the rooted 
theory to the Symanzik effective theory the $a_c m_R(a_c)$ term can be dropped 
from the denominator in Eq.~(\ref{eq:estimate}) since matching
to the taste breaking operators is done at some finite momentum $p \gg \Lambda_{IR}$ which serves in the 
matching loop diagrams as an IR cutoff. 

The bound in Eq.~(\ref{eq:estimate}) is much weaker than needed in the derivation of the SET,
and it implies for infinite volume that the chiral $m\rightarrow 0$ limit
can only be taken after the continuum ($a_f\to 0$) limit.
In Eq.~(\ref{eq:estimate}) the role of $m_R(a_c)$ was to establish the existence of the correct continuum limit
of the full rooted theory on any scale including the far infrared when the volume is infinite~\cite{Shamir:2006nj}.
It follows from~\cite{Bernard:2007ma} that the Symanzik effective theory  is well-defined in the chiral limit,
together with the chiral effective
theory that can be derived from the SET. The requirement that the zero mass
limit for staggered fermions
should be taken only after the continuum limit
is then reproduced by calculations within staggered ChPT \cite{Bernard:2004ab} for certain operators.

To bridge the current work with inherently non-perturbative large volume analysis we follow the procedure just outlined with mass deformed 
analysis at finite cutoff.  What we observe is consistent with chiral symmetry breaking 
of the non-perturbative phase in large volumes. To build the bridge to the results in this paper 
we are interested in a scale-dependent and volume independent renormalized coupling 
in the symmetry breaking phase matching
the scale dependent coupling $g^2_R(a_c)$ presented here. This would leave no room for the $\beta$ function turning zero 
on any scale. This strategy is outlined in more detail in~\cite{Fodor:2015vwa} 
with results of a preliminary implementation.

\section{Numerical simulation}
\label{numericalsimulations}

The details of the simulations are similar to \cite{Fodor:2012td, Fodor:2015baa}. 
In particular we use the staggered fermion action
with 4 steps of stout improvement \cite{Morningstar:2003gk} and stout parameter $\varrho = 0.12$. The bare fermion mass is
set to zero, anti-periodic boundary conditions in all four directions are imposed on the fermions and the gauge field
is periodic. The gauge action is the tree-level improved Symanzik action \cite{Symanzik:1983dc, Luscher:1984xn}.
For integration along the gradient flow we use both the Wilson plaquette and the tree-level improved Symanzik
discretizations. The observable $E(t)$ is discretized as in \cite{Luscher:2010iy}. Hence, in the terminology of 
\cite{Fodor:2014cpa}, we consider the discretizations $WSC$ and $SSC$ for Wilson-flow and tree-level improved
Symanzik-flow, respectively.

\begin{figure}
\begin{center}
\includegraphics[width=7.5cm]{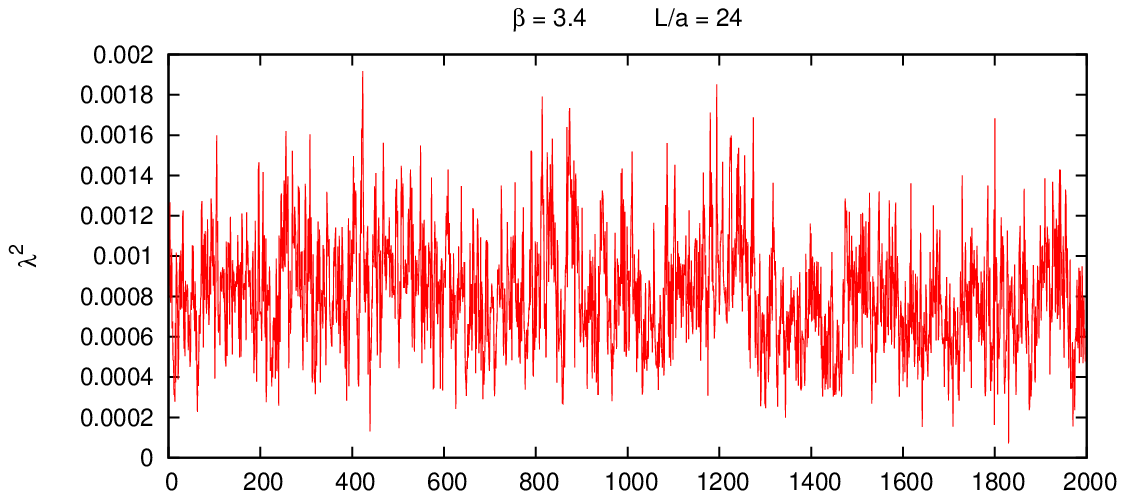} \includegraphics[width=7.5cm]{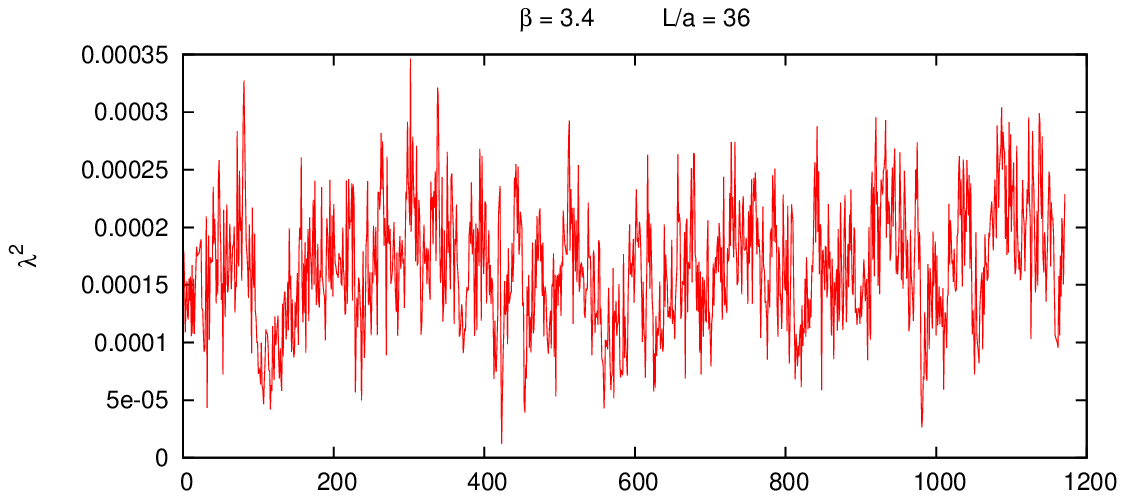}
\includegraphics[width=7.5cm]{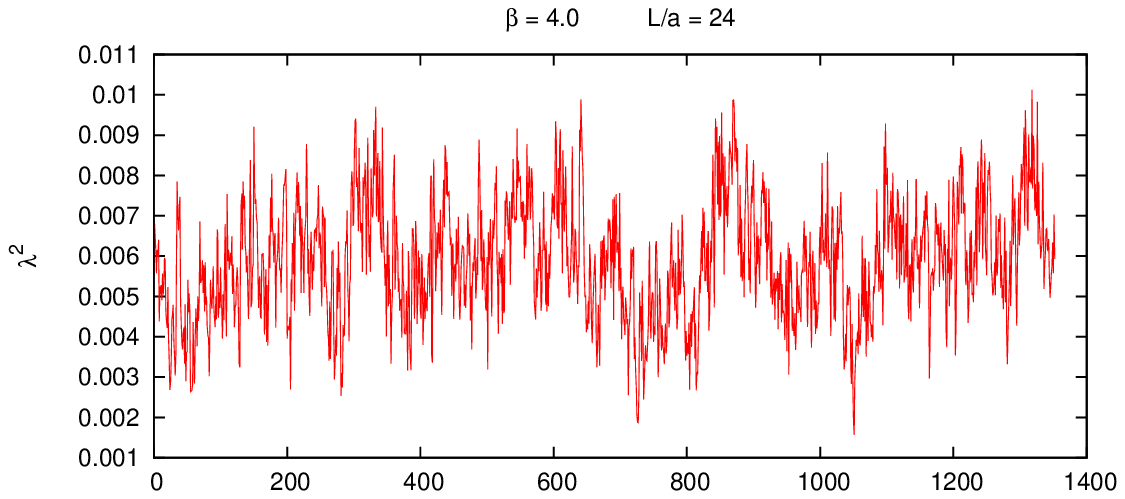} \includegraphics[width=7.5cm]{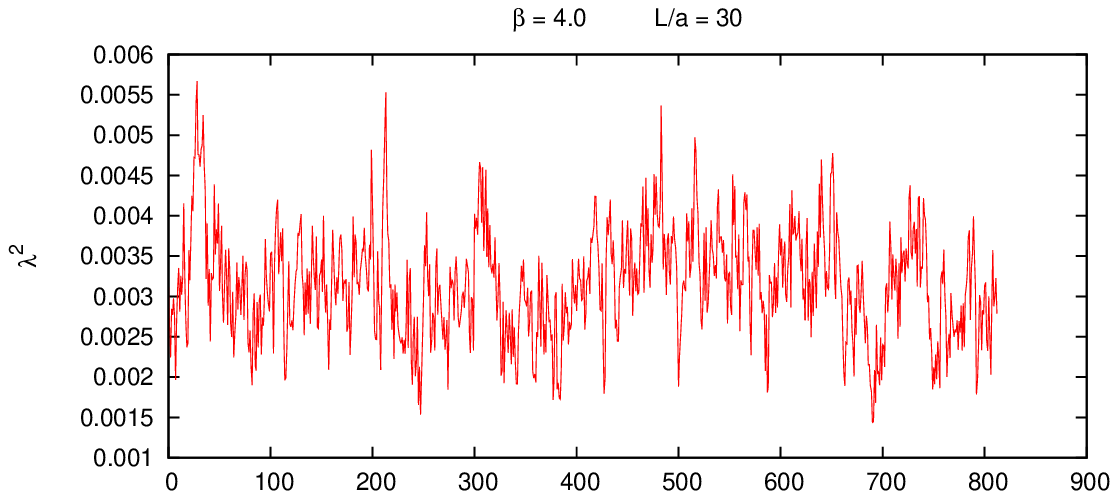}
\end{center}
\caption{Monte-Carlo history of the lowest Dirac eigenvalue, measurements were done for every $10^{th}$ trajectory. The
total number of trajectories are between 8000 and 20000.}
\label{eigenvals}
\end{figure}

As detailed in section \ref{rootedstaggeredformulation} a gap in the Dirac spectrum is needed for the validity
of rooting hence the available physical volume is limited. This translates into the limitation that 
the renormalized coupling cannot be explored above a certain value with a given set of lattice volumes. 
This limitation is however not unique to our running coupling scheme and not even unique to staggered
fermions. All running coupling studies that are directly at the massless limit (by either setting the mass to zero using
staggered or chiral fermions, or tuning $\kappa$ to the massless point $\kappa_c$ using Wilson fermions) 
will be limited to a certain
renormalized coupling range with a given set of lattice volumes.
This is because on a given set of lattice volumes a quite large renormalized coupling can only be achieved by increasing the
bare gauge coupling which in turn will produce small Dirac eigenvalues which in turn will cause the (R)HMC algorithm to 
break down because the condition number of the Dirac operator might be very large on some configurations.

We will see that in our scheme we are able to explore the range
$0 < g_R^2 < 6.5$ which is however quite large and includes the location of the 3-loop and 4-loop fixed point in the $\msbar$
scheme \cite{Vermaseren:1997fq, Pica:2010xq}.

\begin{table}
\begin{center}
\begin{tabular}{|l||l|l|l|l|l|l|l|}
\hline
$L/a\;\;\;\;\beta$& 3.2     &  3.4     &  3.6     &  4.0     &  5.0      &  7.0      &  11.0     \\
\hline
\hline
8                 & 6.90(1) &  5.92(1) &  5.011(8)&  3.58(1) &  1.982(5) &  1.058(3) &  0.547(1) \\
\hline
12                & 7.19(2) &  6.33(1) &  5.44(2) &  4.02(1) &  2.289(5) &  1.220(4) &  0.632(2) \\
\hline
16                & 7.34(2) &  6.47(2) &  5.66(2) &  4.19(2) &  2.410(9) &  1.281(3) &  0.666(3) \\
\hline
18                & 7.41(3) &  6.57(2) &  5.72(4) &  4.31(1) &  2.46(1)  &  1.311(4) &  0.682(2) \\
\hline
20                &         &  6.65(3) &  5.82(2) &  4.34(1) &  2.49(1)  &  1.337(5) &  0.688(1) \\
\hline
24                & 7.69(4) &  6.73(3) &  5.906(9)&  4.45(2) &  2.56(1)  &  1.373(8) &  0.702(3) \\
\hline
30                &         &  6.86(5) &  6.07(7) &  4.59(4) &  2.66(2)  &  1.379(6) &  0.713(4) \\
\hline
36                &         &  7.08(4) &  6.24(3) &  4.65(4) &  2.64(3)  &  1.40(2)  &  0.714(7) \\
\hline
\end{tabular}
\end{center}
\caption{Measured renormalized coupling values in the $SSC$ setup for $c = 7/20$.}
\label{datas}
\end{table}

\begin{table}
\begin{center}
\begin{tabular}{|l||l|l|l|l|l|l|l|}
\hline
$L/a\;\;\;\;\beta$& 3.2     &  3.4     &  3.6     &  4.0     &  5.0      &  7.0      &  11.0     \\
\hline
\hline
8                 & 9.27(1) &  7.76(1) &  6.410(9)  & 4.43(1) &  2.380(5) &  1.247(3) &  0.638(1)   \\
\hline                                                                            
12                & 8.38(2) &  7.29(1) &  6.21(2)   & 4.51(1) &  2.520(6) &  1.328(4) &  0.684(2)   \\
\hline                                                                            
16                & 8.05(2) &  7.04(2) &  6.12(2)   & 4.49(2) &  2.554(9) &  1.349(3) &  0.698(3)   \\
\hline                                                                            
18                & 7.97(3) &  7.03(2) &  6.09(4)   & 4.55(1) &  2.58(1)  &  1.366(4) &  0.708(2)   \\
\hline                                                                            
20                &         &  7.04(3) &  6.13(2)   & 4.54(1) &  2.59(1)  &  1.383(5) &  0.709(1)   \\
\hline                                                                            
24                & 8.02(4) &  7.01(3) &  6.131(9)  & 4.60(2) &  2.63(1)  &  1.406(8) &  0.717(3)   \\
\hline                                                                            
30                &         &  7.04(5) &  6.22(7)   & 4.70(4) &  2.70(2)  &  1.401(6) &  0.723(4)   \\
\hline                                                                            
36                &         &  7.22(4) &  6.35(3)   & 4.72(4) &  2.67(3)  &  1.41(2)  &  0.721(7)   \\
\hline
\end{tabular}
\end{center}
\caption{Measured renormalized coupling values in the $WSC$ setup for $c = 7/20$.}
\label{dataw}
\end{table}

There is also a practical issue related to the rooting procedure. Rooting is implemented by the RHMC algorithm which relies
on the Remez algorithm. The latter is used
for the computation of the coefficients in the partial fraction expansion of the fourth root. A necessary input for the
Remez algorithm is an upper and lower bound on the spectrum of the Dirac operator squared $D^\dagger D$. 
For $m > 0$ a strict lower bound with staggered fermions is $m^2$. However we set $m=0$ and use the anti-periodic
boundary conditions to produce a gap in the spectrum and no strict lower bound is available in this case. Hence we first 
need to measure the lowest and highest Dirac eigenvalues in all runs and then set the lower and upper bounds accordingly
for the subsequent production runs. We found that this procedure is robust and a carefully chosen lower and upper bound 
on the spectrum is not violated in the production runs. Histories of the lowest eigenvalue for various parameters are
shown for illustration in figure \ref{eigenvals}. As expected, increasing $\beta$ leads to a larger lowest eigenvalue and
similarly decreasing the lattice volume also leads to larger lowest eigenvalues.

\begin{figure}
\begin{center}
\includegraphics[width=12cm]{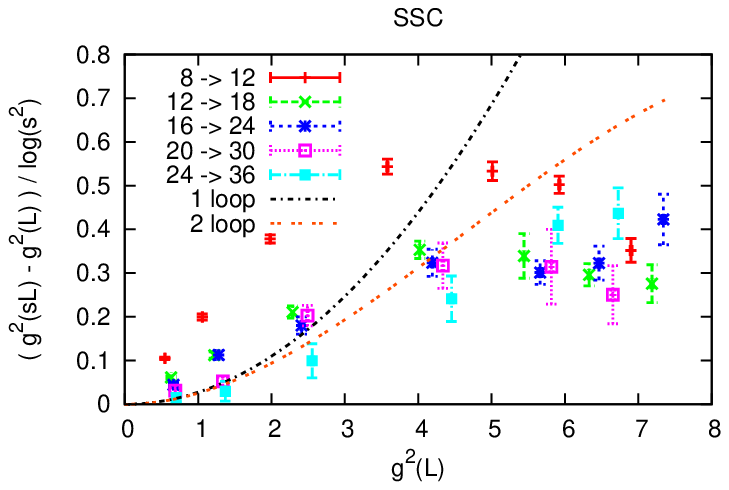} 
\includegraphics[width=12cm]{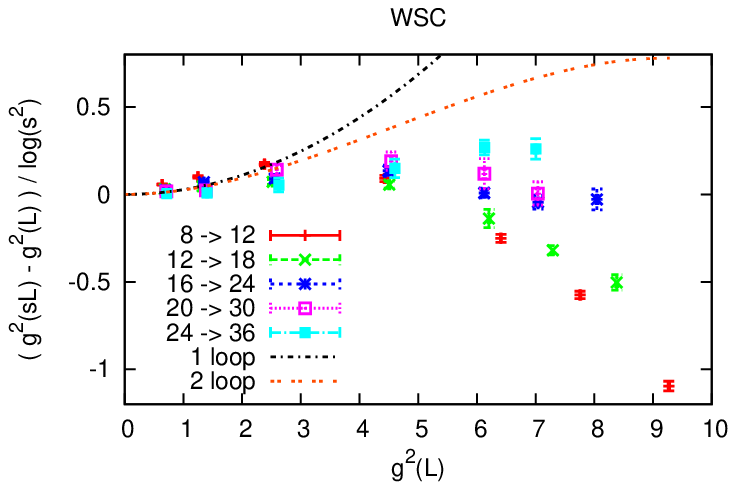}
\end{center}
\caption{Measured discrete $\beta$-function in the $SSC$ (top) and $WSC$ (bottom) discretizations; 
the data correspond to five sets of matched lattice volumes $L \rightarrow sL$ with $s = 3/2$.}
\label{betadata}
\end{figure}


In a lattice setting a convenient and practical method of calculating the running coupling or its $\beta$-function is via step
scaling \cite{Luscher:1991wu, Luscher:1992an}. In this context the finite volume $L$ is increased by a factor $s$
and the change of the coupling, $( g^2(sL) - g^2(L) ) / \log( s^2 )$, is defined as the discrete $\beta$-function.
Note that in this convention asymptotic freedom corresponds to a positive discrete $\beta$-function for small
values of the renormalized coupling.
If the ordinary infinitesimal $\beta$-function of the theory possesses a fixed point, the discrete $\beta$-function will
have a zero as well. Note that as $s\to 1$ the discrete $\beta$-function turns into the infinitesimal variant.
On the lattice the linear size $L$ is easily increased to $sL$ by simply increasing
the volume in lattice units, $L/a \to sL/a$ at fixed bare gauge coupling. In the current work we set $s = 3/2$ and use
volume pairs $8^4 \to 12^4,\; 12^4 \to 18^4,\; 16^4 \to 24^4,\; 20^4 \to 30^4$ and $24^4 \to 36^4$. 
The continuum limit corresponds to $L/a \to \infty$. Hence our data set has 5 pairs of lattice volumes over a range 
of lattice spacings to cover a desired range of renormalized couplings.

The collected number of thermalized unit length trajectories at each bare coupling and 
volume was between 2000 and 20000 depending
on the parameters and every 10$^{th}$ was used for measurements. The acceptance rates were between 65\% and 95\%.
The measured renormalized coupling values are listed in tables \ref{datas} and \ref{dataw} and the resulting discrete
$\beta$-functions are shown in figure \ref{betadata} for the two discretizations we considered, $SSC$ and $WSC$. 

Clearly, at finite lattice spacing, or equivalently at finite lattice volume, the qualitative features of the two
discretizations are quite different. While the discrete $\beta$-function is positive for the $SSC$ setup it turns
negative for the four roughest lattice spacings, i.e. $8^4 \to 12^4$, $12^4 \to 18^4$, $16^4 \to 24^4$ and $20^4 \to
30^4$ for the $WSC$ setup. On the finest lattice spacings, corresponding to $24^4 \to 36^4$, it does stay positive even
in the $WSC$ case, however.
It is important to point out that the observed zeros of the discrete $\beta$-functions of the $WSC$ setup 
for the roughest four lattice spacings are however
such that as the lattice spacing {\em decreases}, the location of the zero {\em increases}.

Let us emphasize that the behavior of the discrete $\beta$-function {\em at finite lattice volume}, 
whether it crosses zero or not, is entirely irrelevant as far as the continuum model is concerned. The measured data at
finite lattice volume need to be continuum extrapolated and zeros of the discrete $\beta$-function may or may not
survive the continuum limit. It will turn out in the next section that in fact the zeros of the $WSC$ setup 
do disappear in the continuum limit while there aren't any zeros to begin with in the $SSC$ setup, 
and the continuum results for the $WSC$ and $SSC$ setups agree, as they should, and show
no sign of a fixed point in the explored coupling range.

\section{Continuum extrapolation}
\label{continuumextrapolation}

The simplest way to perform the continuum extrapolation of our data is to interpolate the renormalized coupling,
$g^2(\beta)$, as a function of the bare coupling $\beta$ at each lattice volume. We choose the interpolating functions
as
\bea
\label{p}
\frac{\beta}{6} - \frac{1}{g^2(\beta)} = \sum_{m=0}^n c_m \left( \frac{6}{\beta} \right)^m\;,
\eea
similarly to \cite{Tekin:2010mm}. The order of the above polynomial is allowed to be $n = 3, 4$ or $5$ for the volumes
$L/a = 8, 12, 16, 18, 24$ and  $n = 3$ or $4$ for the volumes $L/a = 20, 30, 36$. The corresponding degrees of freedom
of the fits are $1, 2$ or $3$ for the first set and $1$ or $2$ for the second set.

Once the parametrized curves $g^2(\beta)$ are obtained for all volumes 
the discrete $\beta$-function $( g^2(sL) - g^2(L) ) / \log( s^2 )$ can be computed for
arbitrary $g^2(L)$ for fixed $L/a$ and $s = 3/2$. Then assuming that corrections are linear in $a^2/L^2$ the continuum
extrapolation can be performed for each $g^2(L)$. 

In \cite{Fodor:2014cpa, Fodor:2014cxa} we calculated the tree-level improvement of our observables in order to have
smaller slopes in the continuum extrapolations. For the $SU(3)$ fundamental model with $N_f = 4$ tree-level improvement
did indeed decrease the slopes over the full considered coupling range, however with $N_f = 8$ we observed in \cite{Fodor:2015baa}
that tree-level improvement only decreased the slopes for small couplings but in fact increased it for larger couplings.
In the current work we observe the same. More precisely for approximately $g^2(L) \lesssim 3.0$ tree-level 
improvement decreased the absolute value of the slope
of the continuum extrapolation but for $g^2(L) \gtrsim 3.0$ it increased it. The reason most probably is the same as for $N_f
= 8$, namely that the large fermion content enhances the fermion loops which are completely absent from the tree-level
calculation and these fermion loops are bound to increase with increasing coupling. For this reason we do not include
tree-level improvement in the current work because the phenomenologically interesting region is in the larger coupling
range, $g^2(L) \sim 6$.

At small values of the renormalized coupling the continuum discrete $\beta$-function can be reliably calculated in continuum
perturbation theory. For the $SU(3)$ sextet model with $N_f = 2$ we have,
\bea
\label{dbeta}
\frac{g^2(sL)-g^2(L)}{\log(s^2)} = b_1 \frac{g^4(L)}{16\pi^2} + \left( b_1^2 \log(s^2) + b_2 \right)
\frac{g^6(L)}{(16\pi^2)^2} + \ldots 
\eea
where $b_1 = 13/3$ and $b_2 = -194/3$. Due to the small volume gauge dynamics mentioned in 
section \ref{thegradientflow} only the first coefficient is the same in our finite volume scheme as above, but
nevertheless for comparison we show both the 1-loop and 2-loop expressions.
The numerical results, after continuum extrapolation, should agree with the perturbative
result for small renormalized coupling and this test is an important cross-check of our procedures. 

Following the above procedure with a fixed polynomial order for each volume for the interpolations one obtains a continuum
result for both the $SSC$ and $WSC$ setups. The interpolations (\ref{p}) are linear in the free parameters hence the
statistical errors are easy to propagate to the final result. Of course one needs to make sure that the continuum
extrapolations are acceptable from a statistical point of view, for example the $\chi^2/dof$ values are not very large,
and one needs to test whether all 5, or perhaps only 4, or perhaps only 3 lattice spacings are in fact in the scaling
region. The more lattice spacings that are useable in the continuum extrapolation, the more reliable the result is.

\section{Systematic error estimate}
\label{systematic}

\begin{figure}
\begin{center}
\includegraphics[width=7.5cm]{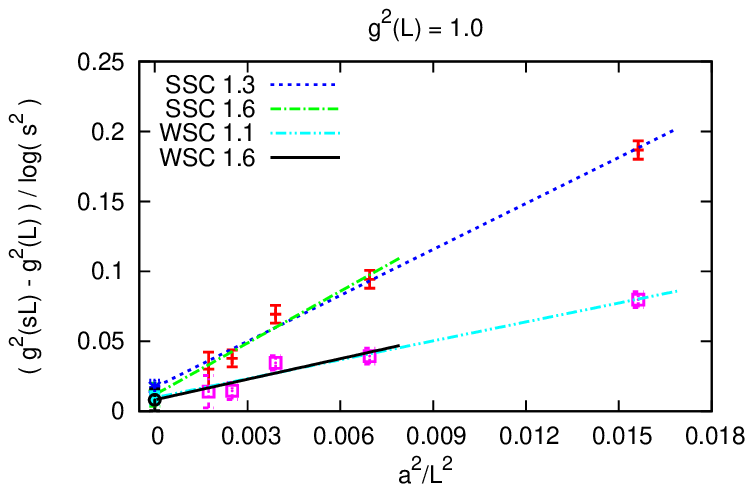} \includegraphics[width=7.5cm]{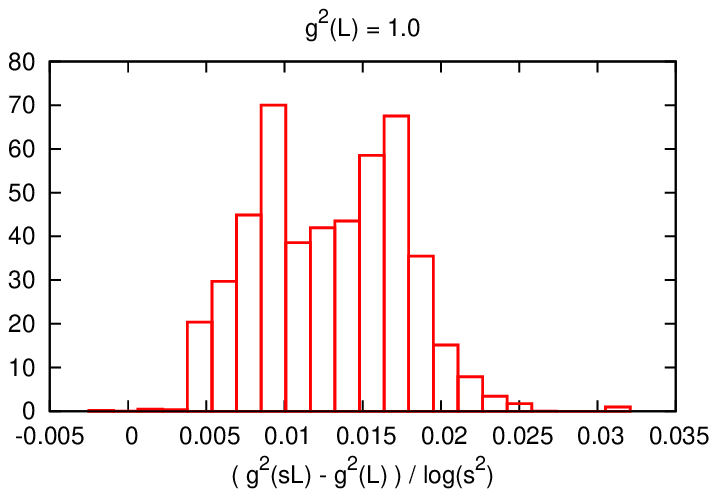} \\
\includegraphics[width=7.5cm]{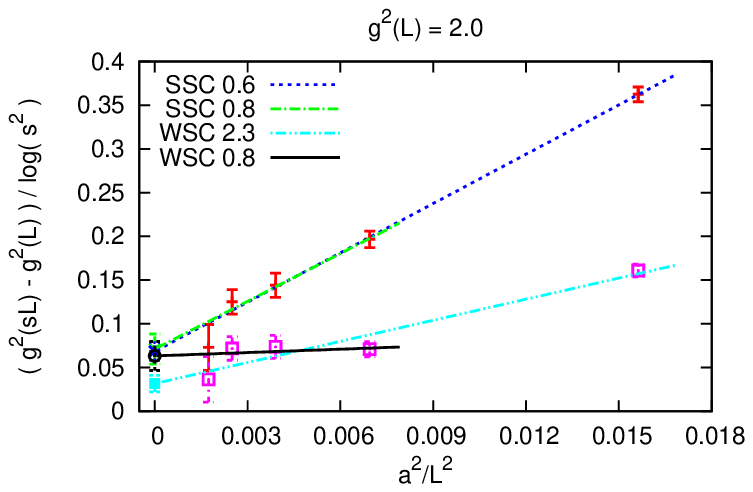} \includegraphics[width=7.5cm]{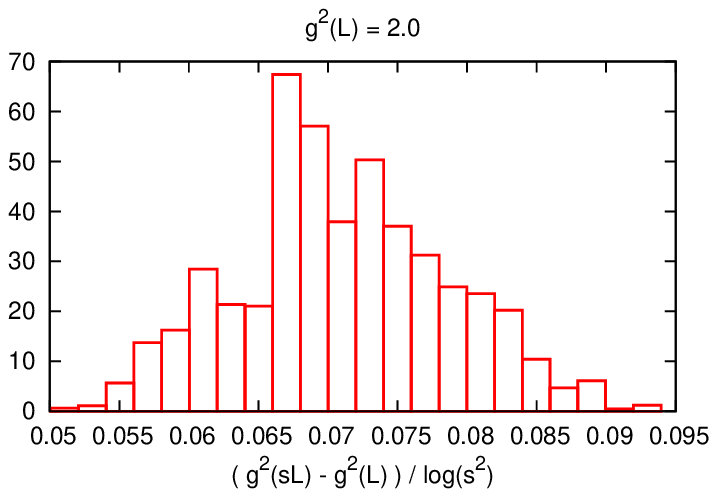} \\
\includegraphics[width=7.5cm]{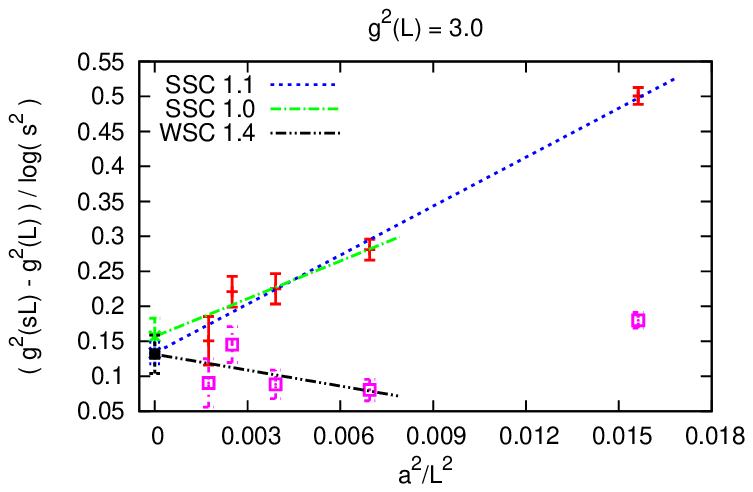} \includegraphics[width=7.5cm]{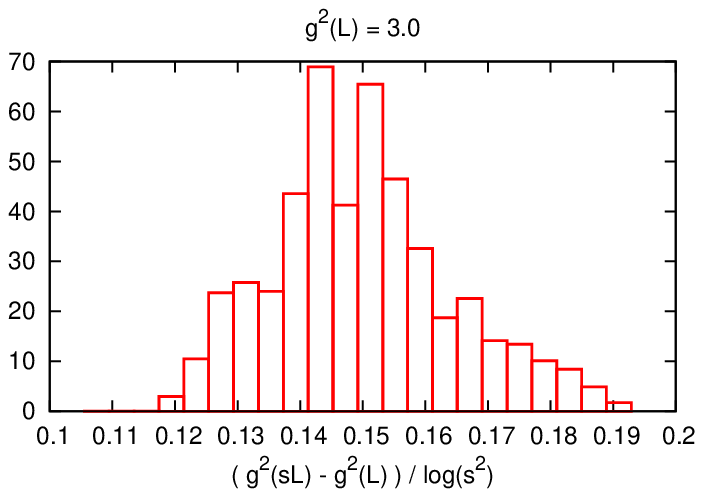} \\
\end{center}
\caption{Right: the weighted histograms of all possible continuum extrapolations used for estimating the systematic
uncertainty for the $SSC$ setup. Left: a representative example of the continuum extrapolations for $g^2(L) = 1.0, 2.0,
3.0$. For comparison we also show a representative example continuum extrapolation in the $WSC$ setup. In each case the
$\chi^2/dof$ of the fit is shown in the legend. If both 5-point and 4-point continuum extrapolations 
can be found around the peak of the histogram, then we show examples from both, otherwise only a 
4-point extrapolation. See text for more details.}
\label{somegsq1}
\end{figure}

\begin{figure}
\begin{center}
\includegraphics[width=7.5cm]{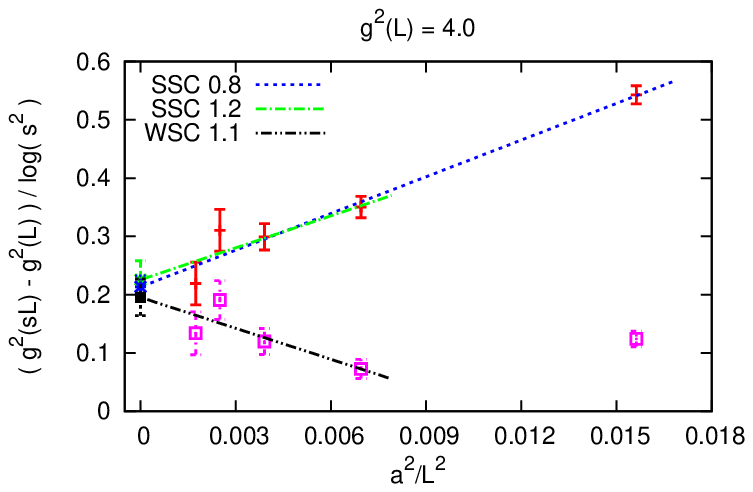} \includegraphics[width=7.5cm]{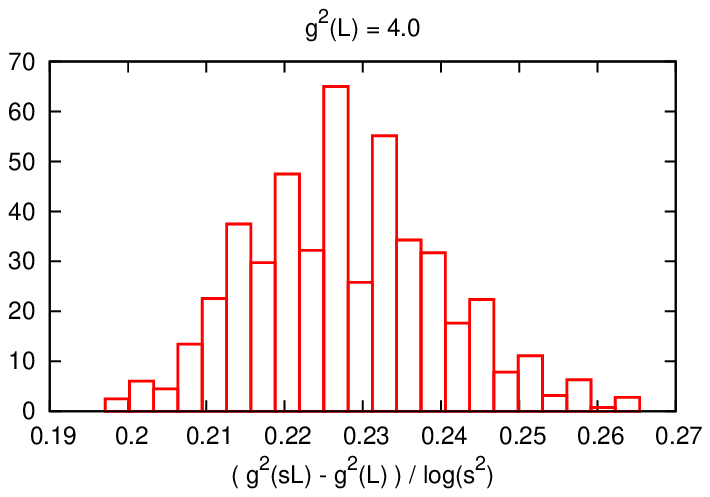} \\
\includegraphics[width=7.5cm]{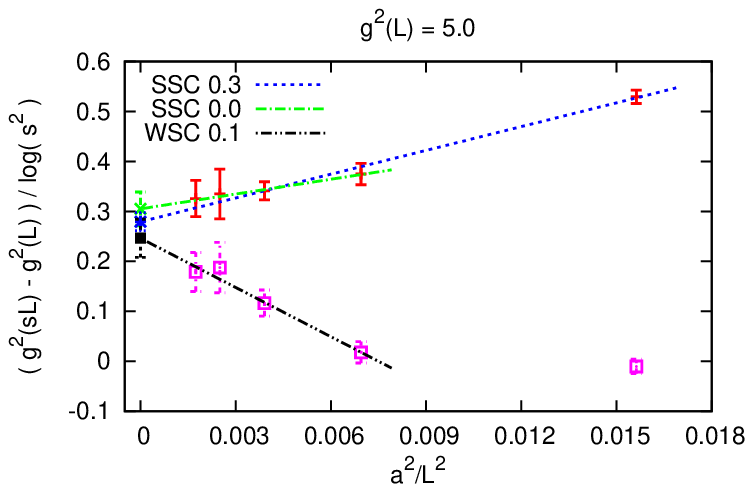} \includegraphics[width=7.5cm]{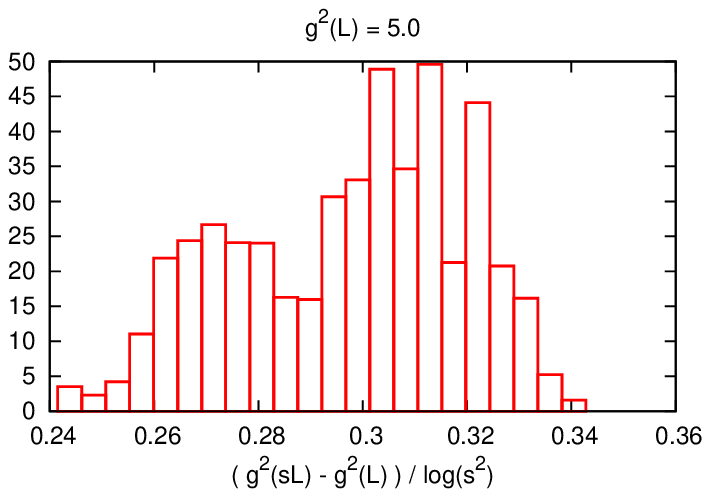} \\
\includegraphics[width=7.5cm]{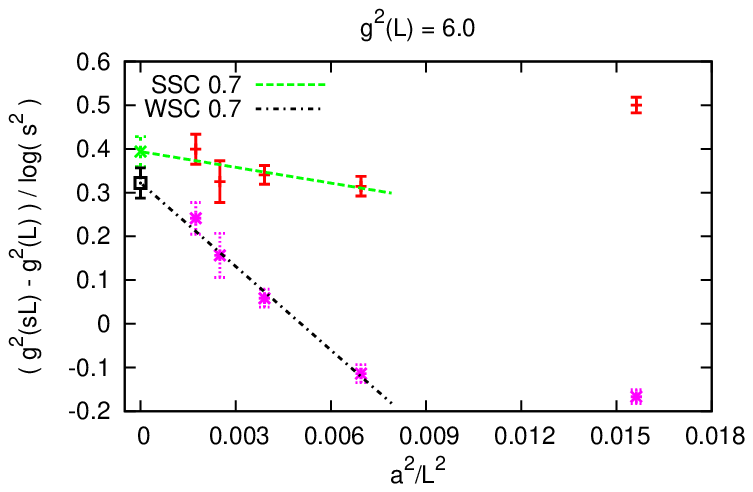} \includegraphics[width=7.5cm]{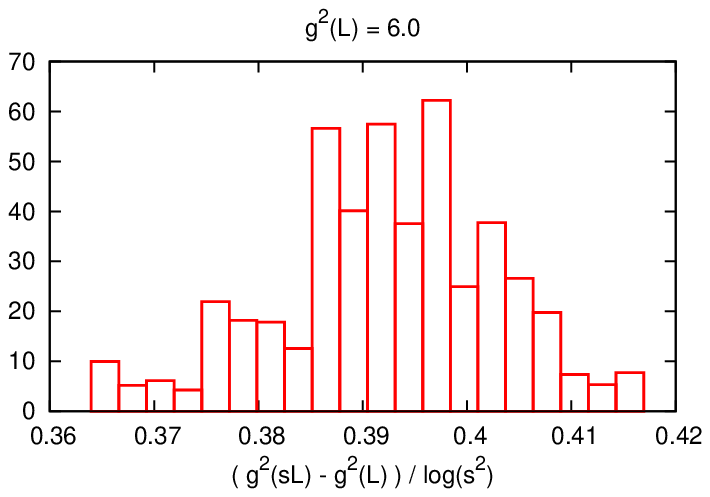} \\
\end{center}
\caption{Right: the weighted histograms of all possible continuum extrapolations used for estimating the systematic
uncertainty for the $SSC$ setup. Left: a representative example of the continuum extrapolations for $g^2(L) = 4.0, 5.0,
6.0$. For comparison we also show a representative example continuum extrapolation in the $WSC$ setup. In each case the
$\chi^2/dof$ of the fit is shown in the legend. If both 5-point and 4-point continuum extrapolations
can be found around the peak of the histogram, then we show examples from both, otherwise only a 
4-point extrapolation. See text for more details.}
\label{somegsq2}
\end{figure}

Apart from the statistical errors we would like to estimate the systematic errors too as precisely as possible. The only
source of systematic error is the continuum extrapolation. However two distinct types of systematic errors are present
in our procedures. One, various polynomial orders can be used for the {\em interpolation} (\ref{p}) 
for each lattice volume and two, one may perform the continuum {\em extrapolation} using 5 or 4 lattice spacings (assuming of
course that all 5 lattice spacings are actually in the scaling region), i.e.~dropping the roughest lattice spacing. As
we discussed in section \ref{numericalsimulations} the rooting trick of the staggered formulation itself
does not introduce unwanted systematic effects.

We will apply the histogram method \cite{Durr:2008zz} in order to estimate the systematic uncertainties.
The polynomial order $n$ for the interpolation (\ref{p}) is allowed to be $n=3,4,5$ for $L/a=8,12,16,18,24$ and $n=3,4$
for $L/a=20,30,36$. All together this leads to $3^5 \cdot 2^3 = 1944$ interpolations and correspondingly to $1944$
continuum results for a given discretization.
Following \cite{Borsanyi:2014jba} a Kolmogorov-Smirnov test is applied to the $1944$ interpolations and only those are
deemed acceptable to which the Kolmogorov-Smirnov test assigns at least a $30\%$ probability. This requirement results
in $240$ and $306$ acceptable interpolations for the $SSC$ and $WSC$ cases, respectively. These all correspond to
continuum extrapolations using 5 lattice spacings.

In order to include the systematic effect coming from performing continuum extrapolations using 4 lattice spacings only,
i.e.~dropping the roughest, $8^4 \to 12^4$, we include such extrapolations too. Using the volumes
$L/a=12,16,18,20,24,30,36$ only with the polynomial orders as above, we have a total number of $3^4 \cdot 2^3 = 648$
interpolations. Out of these the Kolmogorov-Smirnov test allows $240$ and $249$ for the $SSC$ and $WSC$ cases,
respectively\footnote{The fact that the number of allowed interpolations is the same, $240$, for the $SSC$ case for both the
5-point extrapolation and the 4-point extrapolations is purely accidental.}.

Summarizing the above, we have $240 + 240 = 480$ continuum results for the $SSC$ case and $306 + 249 = 555$ continuum
results for the $WSC$ case. In both cases these are binned into a weighted histogram where the weights are given by the
Akaike Information Criterion (AIC) \cite{aic1, aic2, aic3}. We take $68\%$ of the full distribution around the
average to estimate the systematic error. Further details are given in~\cite{Fodor:2015baa} where it is explained how to perform the Kolmogorov-Smirnov test in a running coupling setup and also for the precise definition of the AIC weights.

\section{Final results}

In the final continuum result the statistical and systematic errors are added in quadrature. 
Examples of the weighted histograms for $g^2(L) = 1.0, \ldots, 6.0$ in the $SSC$ setup are shown in the right panels
of figures \ref{somegsq1} and \ref{somegsq2}. For the same renormalized coupling values we show in the left panels 
some representative examples of continuum extrapolations for both the $SSC$ and $WSC$ setups and indicate the
$\chi^2/dof$ values of the fits in the legend.
If all 5 lattice spacings are in the scaling region we include
an example with 5 lattice spacings and also one with 4 lattice spacings. From these plots the following can be inferred.

For approximately $g^2(L) \lesssim 2.5$ all 5 lattice spacings are in the scaling region and the 4-point and 5-point 
continuum extrapolations agree for the $WSC$ setup,
while the same is true for the $SSC$ setup for $g^2(L) \lesssim 5.5$, i.e.~on a much larger range.
Hence for $g^2(L) \gtrsim 2.5$ only the 4-point continuum extrapolations contribute for the $WSC$ setup,
as the 5-point extrapolations are completely suppressed by the AIC weights due to the large $\chi^2$. 
On the other hand for the $SSC$ setup over almost the entire range of renormalized couplings
all 5 lattice spacings are in the scaling regime and the 4-point and 5-point extrapolations agree. For this reason
in the final result we only use the $SSC$ data. However the listed examples in figures \ref{somegsq1} and \ref{somegsq2}
show that the final continuum result using the $WSC$ data actually agrees within errors with the one obtained using the
$SSC$ data. This agreement between two different discretizations is a reassuring consistency check of our procedures,
especially because we have seen in the $WSC$ case in figure \ref{betadata}
that at the smaller lattice volumes the $\beta$-functions did cross zero. 
A remnant of
the small lattice volume $\beta$-functions crossing zero is that for approximately $g^2(L) \gtrsim 5.0$ some of the
$\beta$-function values that are used in the extrapolation are negative. But it is clear from figure \ref{somegsq2} that
the continuum value is positive for both $g^2(L) = 5.0$ and $6.0$ and in fact over the entire range.
The zeros of the small volume $\beta$-functions hence did not survive the continuum
limit and the $WSC$ and $SSC$ final results agree within errors. 

It is worth emphasizing again: in a given discretization the finite (perhaps small) volume discrete $\beta$-functions
can perfectly well cross zero while in another discretization the same thing may not happen. This in itself however is
in no way indicative of the behavior in the continuum as these small volume zeros may disappear in the continuum. The
present model, $SU(3)$ gauge theory with $N_f = 2$ flavors of sextet massless fermions using the $WSC$ and $SSC$ discretizations
serves as an example.

Another cautionary note is in order regarding small volumes. It is clear from figures \ref{somegsq1} and \ref{somegsq2}
that for approximately $g^2(L) \lesssim 5.5$ using only the 3 roughest lattice spacings, 
$8^4 \to 12^4$, $12^4 \to 18^4$ and $16^4 \to 24^4$ would in fact give a continuum result 
compatible with the one including all 5 lattice spacings in the $SSC$
setup. Only at around
$g^2(L) \sim 6.0$ the 3 roughest lattice spacings alone are not usable in a continuum extrapolation. The same, however,
is not the case for the $WSC$ discretization. Already for approximately $g^2(L) \gtrsim 2.5$ the 3-point continuum
extrapolations using only $8^4 \to 12^4$, $12^4 \to 18^4$ and $16^4 \to 24^4$ would result in very high
$\chi^2/dof$ values. If one were to use $8^4 \to 12^4$ and $12^4 \to 18^4$ only as an estimate, this would lead to
a continuum result which is much lower than the reliable 4-point or 5-point continuum extrapolations.
Hence the larger volumes $L/a > 24$ are essential, without these one may obtain 
a much smaller $\beta$-function which actually would be totally unreliable. This
is all the more important in the phenomenologically important larger coupling region $g^2(L) \sim 6$. 

In \cite{annatalks2, annatalks1, annatalks3} 
preliminary results were reported on the sextet model favoring an infrared 
fixed point in the continuum. The scheme used is the same as ours except that the fermions were anti-periodic in
one direction only. The lattice discretization was different, Wilson fermions were used, however the largest lattice
volumes used in the step function were $16^4 \to 24^4$. We speculate the
reported fixed point was a lattice artefact due to the large cut-off effects inherent in using small lattice volumes. This
scenario would be analogous to some extent with our $WSC$ setup and using only our roughest 3 lattice spacings.
Similarly, the inconclusive findings in \cite{Shamir:2008pb, DeGrand:2010na} we speculate are also 
the result of using
small lattice volumes without reaching the scaling regime.

If one were to work with a fixed discretization one of course would not know {\it a priori} how large volumes are needed for a
reliable continuum extrapolation. That is why it is extremely important to consider several discretizations, check that
the lattice spacings are in fact in the scaling region, estimate the
systematic uncertainty coming from the continuum extrapolation reliably, and only trust results in the continuum if they 
agree for the considered discretizations. In our work we have performed this analysis in a fully controlled fashion.

\begin{figure}
\begin{center}
\includegraphics[width=12cm]{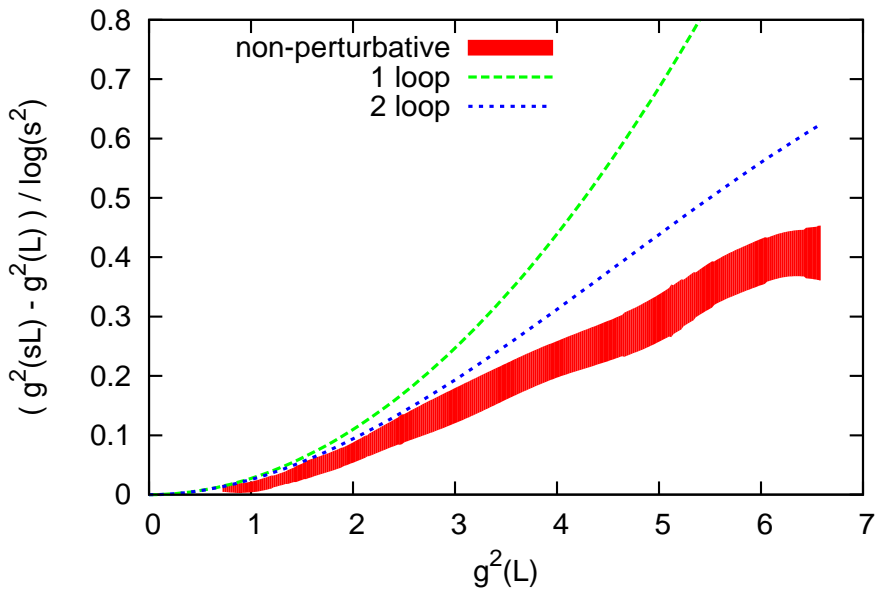} 
\end{center}
\caption{Continuum extrapolated discrete $\beta$-function for $s=3/2$ and $c=7/20$ using the $SSC$ setup.}
\label{final}
\end{figure}

We show the final continuum result in figure \ref{final}. Clearly the $\beta$-function stays positive over the entire
range and is monotonically increasing, and agreement is found for $g^2(L) < 2.5$ between our result and the 2-loop
perturbative result within $1.3 \sigma$.

\section{Conclusion and outlook}

We have studied $SU(3)$ gauge theory coupled to $N_f = 2$ flavors of massless Dirac fermions in the sextet
representation. Our primary motivation was the fact that this model may be a possible realization of a composite Higgs
scenario. The current work is an integral part of our program to understand the infrared dynamics of this model and its
ultimate viability as a building block of Beyond Standard Model physics. Using lattice simulations we have studied the
hadron spectrum and low energy behavior of the model in large volumes in previous work and concluded that 
the model is consistent with spontaneous chiral symmetry breaking. The current work supports this picture in that the running
coupling, when carefully continuum extrapolated, shows no sign of a fixed point in the range $0 < g^2 < 6.5$. This range
in fact includes the zero of the 3-loop and 4-loop $\beta$-function in the $\msbar$ scheme which is $g^2 \simeq 6.28$ and
$g^2 \simeq 5.73$, respectively \cite{Vermaseren:1997fq, Pica:2010xq}. Even though
our scheme is of course different from $\msbar$ and we were only able to explore a limited renormalized coupling range
we are tempted to speculate that the zero of the 3-loop and 4-loop $\beta$-function is a perturbative artefact 
just as the zero of the 2-loop $\beta$-function \cite{Caswell:1974gg, Banks:1981nn}
is a perturbative artefact at $g^2 \simeq 10.58$. The absence of a fixed point and hence
chiral symmetry breaking in the infrared would be consistent with the ladder resummation in the Schwinger-Dyson approach 
predicting non-conformal behavior \cite{Sannino:2004qp, Ryttov:2010iz}.
What we can firmly conclude from our
work is that in the scheme we use, a fixed point is ruled out in the range $0 < g^2 < 6.5$ with high confidence.
In future work we would like to connect the running coupling at some large renormalized coupling value to another scheme
we define in nearly infinite volumes using the flow time as running scale $\mu = 1/\sqrt{8t}$. Since our large volume
simulations are in the chirally broken phase this connection would follow the coupling from the perturbative regime all
the way to the chirally broken regime, ruling out conformal behavior completely.

It is important to emphasize that only non-perturbative lattice calculations are able to reliably answer questions about
the infrared dynamics of non-abelian gauge theories such as our sextet model. These lattice calculations are
nevertheless plagued by systematic uncertainties and reliable results are only possible to obtain if these are fully
controlled. In our work we were able to fully control all systematics leading to reliable continuum results. As far as
the sextet model is concerned, our present work is the first to do so.

The lack of a reliable and fully controlled continuum extrapolation prior to our work made it possible that seemingly
contradictory claims appeared in the literature. These were not contradictory in a sense 
that at finite lattice spacing different discretizations may indeed lead to different conclusions. 
A contradiction only arises if results at finite lattice spacing are interpreted to reflect the properties of the
continuum model. In the continuum of course all correct discretizations should agree and all conclusions should converge.

\acknowledgments

This work was supported by the DOE grant DE-SC0009919,
by the Deutsche Forschungsgemeinschaft
grants SFB-TR 55 and by the NSF under grants 0704171, 0970137, 1318220 and PHY11-25915
and by OTKA under the grant OTKA-NF-104034.
We received major support from an important ALCC Award on the BG/Q Mira
platform of ALCF.
Computations were also carried out at the gpu clusters of Fermilab under our USQCD award, 
on the GPU clusters at the University of Wuppertal, on Juqueen at FZJ and the Eotvos
University in Budapest, Hungary and at the University of California San Diego, USA using the 
CUDA port of the code \cite{Egri:2006zm}.  JK is greatful to Claude Bernard and Maarten Golterman for very helpful discussions 
on their rooting related work. Kalman Szabo and Sandor Katz are  acknowledged for their help and important code development.
KH wishes to thank the Institute for Theoretical Physics and the Albert
Einstein Center for Fundamental Physics at the University
of Bern and the Schweizerischer Nationalfonds for their support.

\end{document}